\documentclass[aps,pre,onecolumn,showpacs,floatfix]{revtex4-1}
\usepackage{graphicx}
\usepackage{dcolumn}
\usepackage{bm}
\usepackage{epsf}
\usepackage{subfigure}
\usepackage{epstopdf}
\usepackage{amsmath}
\DeclareMathOperator\arctanh{arctanh}
\usepackage{amssymb}
\usepackage[cspex,bbgreekl]{mathbbol}
\usepackage{color}
\usepackage{stackrel}
  
\usepackage[utf8]{inputenc}
\usepackage{amsfonts}
\usepackage{times,fullpage}
\usepackage{comment}
\usepackage{array} 
\usepackage{verbatim}

\newcommand{\be}{\begin{equation}}
\newcommand{\ee}{\end{equation}}
\newcommand{\ba}{\begin{eqnarray}}
\newcommand{\ea}{\end{eqnarray}}
\newcommand{\g}{\chi}

\newcommand{\nwc}{\newcommand}
\nwc{\vs}{\vspace}
\nwc{\hs}{\hspace}
\nwc{\la}{\langle}
\nwc{\ra}{\rangle}
\nwc{\nn}{\nonumber}
\nwc{\Ra}{\Rightarrow}
\nwc{\wt}{\widetilde}
\nwc{\lw}{\linewidth}
\nwc{\ft}{\frametitle}
\nwc{\ben}{\begin{enumerate}}
\nwc{\een}{\end{enumerate}}
\nwc{\bit}{\begin{itemize}}
\nwc{\eit}{\end{itemize}}
\nwc{\dg}{\dagger}
\nwc{\mA}{\mathcal A}
\nwc{\Tr}[1]{\underset{#1}{\mbox{Tr}}~}
\nwc{\pd}[2]{\frac{\partial #1}{\partial #2}}
\nwc{\ppd}[2]{\frac{\partial^2 #1}{\partial #2^2}}
\nwc{\fd}[2]{\frac{\delta #1}{\delta #2}}
\nwc{\pr}[2]{K(i_{#1},\alpha_{#1}|i_{#2},\alpha_{#2})}
\nwc{\av}[1]{\left< #1\right>}
\nwc{\rlh}[2]{\underset{#2}{\stackrel{#1}{\rightleftharpoons}}}

\begin{document}

\title{Kinetics and thermodynamics of reversible polymerization in closed systems}

\author{Sourabh Lahiri$^1$, Yang Wang$^1$, Massimiliano Esposito$^2$, David Lacoste$^1$}
\affiliation{$^1$ Laboratoire de Physico-Chimie Th\'eorique, UMR CNRS Gulliver 7083, ESPCI - 10 rue Vauquelin, F-75231 Paris, France}
\affiliation{$^2$ Complex Systems and Statistical Mechanics, University of Luxembourg, L-1511 Luxembourg, Luxembourg}

\date{\today}

\begin{abstract}
Motivated by a recent work on the metabolism of carbohydrates in bacteria, we study 
the kinetics and thermodynamics of two classic models for reversible polymerization,
one preserving the total polymer concentration and the other one not.
The chemical kinetics is described by rate equations following the mass-action law.
We consider a closed system and nonequilibrium initial conditions and show that the 
system dynamically evolves towards equilibrium where detailed balance is satisfied. 
The entropy production during this process can be expressed as the time derivative of a Lyapunov function. 
When the solvent is not included in the description and the dynamics conserves the 
total concentration of polymer, the Lyapunov function can be expressed as a Kullback-Leibler 
divergence between the nonequilibrium and the equilibrium polymer length distribution. 
The same result holds true when the solvent is explicitly included in the description and 
the solution is assumed dilute, whether or not the total polymer concentration is conserved. 
Furthermore, in this case a consistent nonequilibrium thermodynamic formulation can be established 
and the out-of-equilibrium thermodynamic enthalpy, entropy and free energy can be identified.    
Such a framework is useful in complementing standard kinetics studies with the
 dynamical evolution of thermodynamic quantities during polymerization.
\end{abstract}


\maketitle
\section{Introduction}


{The processes of aggregation or polymerization are ubiquitous in Nature, 
for instance they are present in the polymerization of proteins, the coagulation of blood, or even in the formation of stars. They are 
often modeled using the classic coagulation equation derived by Smoluchowsky \cite{Smoluchowsky1916,smoluchowsky1916a}.
In the forties, P. Flory \cite{Flory1936,Flory1944} developed his own approach for  
reactive polymers, with an emphasis on their thermodynamic properties and on their most likely (equilibrium) 
size distribution. At this time, only irreversible polymerization was considered. 
The first study of the kinetics of reversible polymerization combining aggregation and fragmentation processes
was carried out by Tobolsky et al. \cite{Blatz1945}.}

{In the seventies, reversible polymerization became a central topic in studies on    
 the association of amino acids into peptides and on the self-assembly of actin. 
The thermodynamics of assembly of these polymers forms the topic of the now classical 
treaty by F. Oosawa and S. Asakura \cite{Oosawa}. 
At about the same time, T. L. Hill made many groundbreaking contributions to non-equilibrium statistical 
physics and thermodynamics, 
which allowed him to describe not only
the self-assembly of biopolymers like actin and microtubules, 
but also to address much more complex questions such that of free energy transduction by biopolymers 
and complex chemical networks \cite{Hill1980}.
In the eighties, R. J. Cohen and G. B. Benedek \cite{Cohen1982} revisited the work of Flory and Stockmayer, by showing 
that the Flory polymer length distribution is obtained under the assumption 
of equal free energies of bond formation for all bonds of the same type. They also showed that 
the kinetically evolving polymer distribution does not have the Flory form in general, and they analyzed 
the irreversible kinetics of the sol-gel transition.   
}

More recently, the specific conditions on the kernels of 
aggregation and fragmentation, for which equilibrium solutions 
of the Flory type exist have been analyzed \cite{Bak1979,Dongen1984,Vigil2009}.
When these conditions are not met, reversible polymerization models admit interesting 
nonequilibrium phase transitions which are beginning to be investigated \cite{Kaprivsky2008, Krapivsky2010_vola}.
 
{The kinetic rate equations of reversible polymerization have broad applications. 
For instance, in studies on the origin of life, these equations 
describe the appearance of long polymer chains in the primordial soup \cite{Braun2013}.}
These equations are also used to describe the formation of protein clusters in membranes 
\cite{Turner2005, Foret2012, Zapperi2014} or the self-assembly of carbohydrates (also called 
glycans) \cite{Kartal2011}. In the latter case, a very large repertoire of polymer structures 
and enzymes are involved in the synthesis and degradation of these polymers. 
Since it is hardly possible to model all the involved chemical reactions, the authors of this work, 
Kartal et al., introduced a statistical approach to explain experiments which they have performed using 
mixtures of such polymers with the appropriate enzymes. 
Their study underlines the importance of entropy as a driving force in the dynamics 
of these polymers: under its action a monodisperse solution of such biopolymers, which 
is placed in a closed reactor with the appropriate enzymes, typically admits an exponential 
distribution of polymer length as equilibrium distribution, in agreement with maximum 
entropy arguments used by P. Flory \cite{Flory1936, Flory1944}. 

This recent work of Kartal et al. \cite{Kartal2011}, motivated us to construct appropriate 
dynamics which converge on long times towards such equilibrium distributions.
In order to complement this with an analysis of the time evolution of thermodynamic quantities, 
we rely on stochastic thermodynamics (for general 
reviews see \cite{Seifert2012, Ritort2008_vol137, Jarzynski2011_vol2, EspVDBRev2014}). 
While this recent branch of thermodynamics has been used extensively in 
the literature for chemical reaction networks \cite{Schnakenberg1976, NicolisVdB84, 
VanDenBroeckST86, Gaspard2004_vol120, SeifertSchmiedlJCP07, QianPR12a, QianPR12b, 
Muy2013} and copolymerization processes \cite{Andrieux2008_vol105} at the level of the 
stochastic chemical master equation, its application to the level of mean-field 
kinetic rate equations is more recent \cite{Polettini2014}.

In this paper, we precisely use this level of description based on mean-field rate equations. 
Implicitly, we assume reaction-limited polymerization. 
Naturally, if the reactions are too fast or the mobility of the polymers too low, a 
mean-field approach may not be sufficient and diffusion processes should be accounted for. 
We focus on two main models of reversible polymerization which 
reproduce the equilibrium distributions found in Ref. \cite{Kartal2011}: 
In the first one, there is only one conservation law (the total number of monomers) 
while in the second one, there are two (the total number of monomers and of polymers). 

The plan of the paper is as follows. In section \ref{sec:General thermo A}, we study reversible polymerization 
using general rate equations compatible with one conservation law (the total number of monomers).
This is done first at the one-fluid level for which there is no solvent, 
and then at the two-fluid level, for which there is a solvent. 
In section \ref{sec:Application string model}, we apply this general framework to a specific
model called String model, in which the rates of aggregation and fragmentation are constant.
In section \ref{sec:General 2 conservation laws}, we extend the previous case of reversible 
polymerization with a single conservation law to the case where there are two conservation 
laws, namely the total number of monomers and of polymers.
In section \ref{sec:DPE1-DPE2}, inspired by Ref. \cite{Kartal2011}, we study two specific examples of reversible 
polymerization with two (resp. three) conservation laws, namely the kinetics of glucanotransferases DPE1 (resp. DPE2). 
For both cases, we construct the dynamics which converge towards the equilibrium distributions found in 
Ref. \cite{Kartal2011}, and we discuss their properties from the standpoint of nonequilibrium thermodynamics. 

\section{Reversible polymerization with one conservation law}
\label{sec:General thermo A}

We consider a reversible polymerization process made of the following elementary reactions
\begin{equation}
[n]+[m] \stackrel[k_{nm}^-]{k_{nm}^+}{\rightleftharpoons} [n+m],
\label{reversible poly general n}
\end{equation}
{where the forward and backward reaction rates, namely $k_{nm}^-$ and $k_{nm}^+$, are functions 
of the polymer lengths $n$ and $m$, which are strictly positive and symmetric under a permutation of $n$ and $m$}.
We denote by $c_l$ the concentration of polymers of length $l$. The evolution of this quantity is ruled by 
\be \label{Dyn}
\dot{c}_l = \frac{1}{2} \sum_{n+m=l>1} ( k_{nm}^+ c_n c_m - k_{nm}^- c_{n+m} ) 
- \sum_{n=1}^{\infty} ( k_{ln}^+ c_l c_n - k_{ln}^- c_{l+n} ), 
\ee
which preserves the total concentration of monomers (i.e. the concentration that one would get if all the polymers were 
broken into monomers) $M=\sum_{l=1}^{\infty} l c_l$, but not the total concentration of polymers $c=\sum_{l=1}^{\infty} c_l$. 
Therefore in this case, there is only one conservation law, that of $M$, and one can assume $M=1$.
{Note also that Eq. \ref{Dyn} generalizes the Becker-D\"oring equations which 
describe the dynamic evolution of clusters that gain or loose only one unit at a time \cite{Becker1935,Penrose1986}} 

Assuming that the reactions (\ref{reversible poly general n}) can be treated as elementary, 
the entropy production rate is given by \cite{PrigogineKondepudi, GrootMazur}   
\be \label{EPChem}
\Sigma = \frac{R}{2} \sum_{n,m} (k_{nm}^+ c_n c_m - k_{nm}^- c_{n+m}) \ln \frac{k_{nm}^+ c_n c_m}{k_{nm}^- c_{n+m}} \geq 0,
\ee
where $R$ is the gas constant. The entropy production rate vanishes when the system reaches equilibrium, 
i.e. when and only when detailed balance is satisfied:
\begin{eqnarray}
k_{nm}^+ c^{eq}_n c^{eq}_m = k_{nm}^- c^{eq}_{n+m}. \label{DB}
\end{eqnarray} 

Using the inequality $\ln x \leq x-1$, which holds for all $x>0$, one easily proves that the 
following quantity is non-negative, convex and vanishes only at equilibrium
\begin{eqnarray}
L \equiv R \sum_{l} c_l \ln \frac{c_l}{c_l^{eq}}- R (c-c^{eq}) \geq 0. \label{def:G}
\end{eqnarray}
Indeed, by taking the time derivative of $L(t)$ and using the definition of $c$, 
one obtains $dL/dt=R \sum_{l} \dot{c}_l \log c_l/c_l^{eq}$. 
Now using (\ref{Dyn}), we obtain two terms. The first term is 
\be
\frac{R}{2} \sum_l \sum_{n+m=l} (k_{nm}^+ c_n c_m - k_{nm}^- c_{n+m}) \ln \frac{c_l}{c_l^{eq}} 
= \frac{R}{2} \sum_{n,m} (k_{nm}^+ c_n c_m - k_{nm}^- c_{n+m}) \ln \frac{c_{n+m}}{c_{n+m}^{eq}}, \nonumber
\ee
while the second one is
\be
- R\sum_l \sum_n (k_{nl}^+ c_n c_l - k_{nl}^- c_{n+l}) \ln \frac{c_l}{c_l^{eq}} 
= -\frac{R}{2} \sum_{n,m} (k_{nm}^+ c_n c_m - k_{nm}^- c_{n+m}) \ln \frac{c_n c_m}{c_n^{eq} c_m^{eq}} . \nonumber
\ee
By adding these two contributions and using the detailed balance condition of (\ref{DB}) as well as (\ref{EPChem}), 
one finds that 
\be
\Sigma = -R \sum_{l} \dot{c}_l \ln \frac{c_l}{c_l^{eq}} = - \frac{d}{dt} L \geq 0. \label{extensive 2nd law}
\ee
This proves that $L$ is a Lyapunov function, i.e. a non-negative, monotonically decreasing function, which vanishes 
at and only at equilibrium. The existence of such a function implies that the dynamics will always relax to a unique 
equilibrium state. 
{In \cite{Shapiro1965}, the authors show that for a chemical system containing a finite number of homogeneous phases, a Gibbs free energy function exists that is minimum at equilibrium. In the context of reacting polymers, a similar free energy function has been derived in \cite{Bak1979}. In view of the non-increasing property of this function, the authors have coined the term ``F-Theorem''.} 
We discuss below its relation to the more usual H theorem.

\subsection{One-fluid version}

Until now, our model was exclusively expressed in terms of concentrations and could be used to study non-chemical 
dynamics such as population dynamics. We now introduce the one fluid model, where the solvent is not described either 
by choice or because it is absent. The molar fractions of the polymers of length $l$, with $l\ge 1$, are 
$x_l=c_l/c$ while the Lyapunov function is
\be
L=R c \sum_{l\ge 1} x_l  \ln \frac{x_l}{x_l^{eq}} + R c \ln \frac{c}{c_{eq}}- R (c-c^{eq}) \geq 0. \label{def:G2}
\ee
It is important to note that this Lyapunov function is in general distinct from the relative entropy or Kullback-Leibler 
divergence between the distribution $x_l$ and $x_l^{eq}$, which represents only the first term in Eq. (\ref{def:G}). 
The reason for this difference is that $c_l$, contrary to $x_l$, cannot 
always be interpreted as a probability since its norm is not always conserved. 
The two quantities become however equivalent (i.e. they only differ by a constant) 
when the total concentration of polymers $c(t)$ is constant in time. 
If furthermore -$\sum_l x_l \ln x_l^{eq}$ is constant in time (the meaning of this assumption 
will become clear in the two fluid model), then the negative of the Shannon entropy (which is related to the classic 
notion of free energy of mixing introduced in \cite{Flory1936,Flory1944} as explained in \cite{Ben2006})  
\be
S_{Sh}=- R\sum_{l\ge 1} x_l \ln x_l, \label{Real-Shannon}
\ee
becomes a Lyapunov function and (\ref{extensive 2nd law}) reduces to the famous Boltzmann ``H theorem''. 

\subsection{Two-fluid version}\label{subsec:1conser_2fluids}

In order to make contact with thermodynamics, we now introduce the two fluid model which {includes} the solvent 
explicitly in the list of chemical species. Then, the molar fraction of the polymer of length $l$ becomes 
$y_l(t)= c_l(t)/C(t)$, which importantly is now defined with respect to the total concentration of all species 
including the (time-independent) solvent concentration $c_0$ (water for instance): $C(t)=\sum_{l\ge 0}c_l(t)=c(t)+c_0$, 
where $c(t)=\sum_{l \ge 1} c_l(t)$. 
If there is no solvent, $c_0=0$ and one recovers the molar fraction of the previous section which was denoted by $x_l$.
In dilute solution, since $y_0$ is very close to one and the other $y_l$ are much smaller, $C(t)$ becomes almost constant: 
$C \approx c_0$. The chemical potential of a polymer of length $l$ in a dilute solution is defined by $\mu_l = \mu^0_l + RT \ln y_l$, 
where $\mu^0_l=h^0_l-Ts^0_l$ is the standard reference chemical potential and $h^0_l$ and $s^0_l$ are the standard 
enthalpy and entropy respectively. We restrict ourselves here to ideal solutions, and by this we assume that this form of chemical potential applies not only to the polymers ($l\neq 0$) but also to the solvent. An interesting study of 
the effect of non-ideality on the time evolution thermodynamic quantities during reversible polymerization
can be found in Ref.~\cite{Stier2006}. 

Let us define the intensive enthalpy function as
\be \label{EnthalpyFct}
H = \sum_{l\ge 0} y_l h_l^0,
\ee
and the entropy function of this two-fluid model as
\be \label{Shannon}
S = \sum_{l\ge 0} y_l (s_l^0 - R \ln{y_l}).
\ee
Their extensive counterparts are ${\cal H}=C H$ and ${\cal S}=C S$. In Eq. (\ref{Shannon}), the first term proportional 
to $s_l^0$ therefore represents the entropic contribution due to the disorder in the internal degrees of freedom of each polymer, 
while the second term represents the nonequilibrium entropy in the distribution of the variables $y_l$ \cite{Gaspard2004_vol120}. 
Let us introduce the intensive free enthalpy
\be
G=H-TS = \sum_{l \ge 0} y_l \mu_l,
\ee
where we used the definition of the chemical potential in the last equality, and its extensive counterpart ${\cal G}=CG$.

Since the change of chemical potential associated to each reaction must vanish at equilibrium, i.e. 
$\Delta \mu = \mu_{n+m} - \mu_n - \mu_m=0$, using the definition of the chemical potential, we find that   
\begin{eqnarray}
RT \ln \frac{y^{eq}_{n+m}}{y^{eq}_n y^{eq}_m} = - (\mu^0_{n+m}-\mu^0_n-\mu^0_m). \label{mass action}
\end{eqnarray}
Combining (\ref{mass action}) with (\ref{DB}), we obtain that the kinetic constants must satisfy local detailed balance
\begin{eqnarray} 
RT \ln \frac{k_{nm}^+C^{eq}}{k_{nm}^-}= - (\mu^0_{n+m}-\mu^0_n-\mu^0_m), \label{LDB}
\end{eqnarray}
where $C^{eq}$ denotes the equilibrium value taken by $C$. 
We note that since the chemical reactions do not involve the solvent, 
we formally define the rate constants with any zero subscript ($n$ or $m$) to be zero. 
Naturally, Eq. \eqref{LDB} is not applicable in this case. 
{The validity of Eq (13) relies on two main assumptions: the first one is that of dilute solutions, 
while the second one is the ideality of the heat bath. The latter assumption means that there are no hidden degrees of freedom
 which can dissipate energy in the chemical reactions under consideration, which implies in particular that these reactions must be elementary.}

Using the definitions of the enthalpy and entropy functions, we find that
\be \label{Eflow}
\frac{d{\cal H}}{dt}  =  \frac{1}{2} \sum_{n,m} (k_{nm}^+ c_n c_m - k_{nm}^- c_{n+m}) \big(h_{n+m}^0-h_n^0-h_m^0 \big) 
\ee
and
\be \label{ShannonDerivative}
\frac{d{\cal S}}{dt}  =  \frac{1}{2} \sum_{n,m} (k_{nm}^+ c_n c_m - k_{nm}^- c_{n+m}) \big(s_{n+m}^0-s_n^0-s_m^0 + R \ln \frac{y_n y_m}{y_{n+m}} \big).
\ee
As a result, one finds that
\be
\frac{d{\cal G}}{dt}  =  \frac{RT}{2} \sum_{n,m} (k_{nm}^+ c_n c_m - k_{nm}^- c_{n+m}) 
\ln \frac{y_{n+m} y^{eq}_n y^{eq}_m}{y^{eq}_{n+m} y_n y_m} = RT \sum_l {\dot c}_l \ln \frac{y_l}{y^{eq}_l},
\ee
where we used Eq.~\eqref{mass action} to obtain the last equality. 
By including the solvent in the sum, as the term corresponding to $l=0$, we have $\sum_{l \ge 0} y_l=1$ and 
therefore $\sum_{l \ge 0} {\dot y}_l=0$; thus we can rewrite the above equation as
\be
\label{newdGdt}
\frac{d{\cal G}}{dt}  = RT \frac{d}{dt} \bigg( C \sum_{l \ge 0} y_l \ln \frac{y_l}{y^{eq}_l} \bigg).
\ee

The entropy production (\ref{EPChem}), using (\ref{LDB}) and the chemical potential definition, may be written as
\be \label{EPChemBis}
T \Sigma = -\frac{1}{2} \sum_{n,m} (k_{nm}^+ c_n c_m - k_{nm}^- c_{n+m}) 
\big(\mu_{n+m}-\mu_n-\mu_m - RT \ln \frac{C}{C^{eq}} \big) \geq 0.
\ee
Using Eq.~\eqref{Eflow} and Eq.~\eqref{ShannonDerivative}, it can be rewritten as
\be \label{EntBalance}
T \Sigma = -\frac{d}{dt} \bigg[ {\cal H}-T{\cal S} -RT C \ln \frac{C}{C^{eq}} - R T \big( C^{eq}-C \big) \bigg] \geq 0.
\ee
The first term is the heat flow, the second the entropy change and the third and fourth terms represent a contribution 
due to the change in the total concentration.
It is important to note that within the two-fluid model with ideal solutions, since $C(t)$ is essentially constant, 
these latter two contributions are negligible. Neglecting these terms, the entropy production can be expressed as the change in free 
energy which is also equal to a change in Kulback-Leibler divergence between the nonequilibrium and the equilibrium polymer distribution
\be \label{EntBalance2}
T \Sigma = -\frac{d{\cal G}}{dt}  = - R T \frac{d}{dt} \big( C \sum_{l \ge 0} y_l  \ln \frac{y_l}{y_l^{eq}} ) \geq 0.
\ee
We finally note that when all the polymerization reactions are neutral from a standard chemical potential standpoint 
i.e. when $\mu^0_{n+m}=\mu^0_n+\mu^0_m$ for all $n,m$, one has that
\be
\frac{d{\cal G}}{dt}  = RT \frac{d}{dt} \big( \sum_{l \ge 0} c_l \ln y_l \big)= RT \frac{d}{dt} \big( C \sum_{l \ge 0} y_l \ln y_l \big),
\ee
which together with Eq.~\eqref{newdGdt} implies $d_t (C \sum_l y_l \ln y_l^{eq})=0$.
Since $C$ can be assumed constant, the Lyapunov function can be expressed only in terms 
of the Shannon entropy constructed from $y_l$ instead of the full KL divergence.
The dynamics can then be compared to a Boltzmann equation where the relaxation to equilibrium is purely driven by the maximization 
of the Shannon entropy, as in the H-Theorem.

\section{Application to the String model}
\label{sec:Application string model}

As a simple realisation of the reversible polymerization given by Eq. (\ref{reversible poly general n}), 
we now consider the String model which assumes constant rates of aggregation and fragmentation, 
independent of the length of the reacting polymers. Following \cite{Krapivsky2010_vola}, we choose 
$k_{nm}^+=2$ and $k_{nm}^-=2 \lambda$. 
From Eq. \eqref{Dyn}, the dynamics follows 
\be
\dot{c}_l=\sum_{i+j=l} c_i c_j + 2 \lambda \sum_{j>l} c_j - 2 c_l c - \lambda (l-1) c_l,
\label{evolution-string-model}
\ee
where we used the fact that $(l-1)$ combinations of $i$ and $j$ satisfy the relation $i+j=l$.
The detailed balance condition defining equilibrium implies that $c_i^{eq} c_j^{eq}= \lambda c_{i+j}^{eq}$, 
which admits one parameter solutions of the form $c_l^{eq}=\lambda \beta^l$. 
Assuming the total monomer concentration to be $M=\sum_l l c_l=1$, one finds
\be
\beta=1+\frac{\lambda}{2} - \sqrt{\lambda+\frac{\lambda^2}{4}},
\label{beta}
\ee
since the solution $c_l$ must decay at large $l$.

\subsection{One-fluid version of the String model}

\begin{figure}[!h] 
\centering
\includegraphics[width=0.4\linewidth]{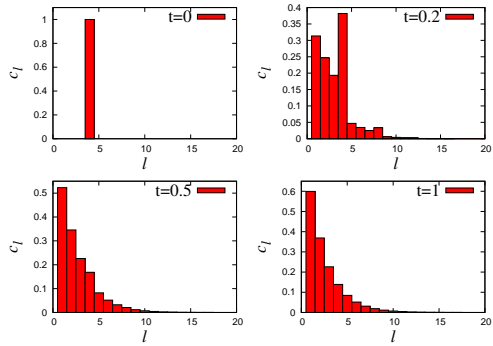}
\caption{
Evolution of the size distribution of polymers $c_l$ for the String model at different times $t=0$, $t=0.2$, 
$t=0.5$ and $t=1$ starting with a monodisperse distribution with characteristic length $l=4$ at time $t=0$.
The transition rates are constant and correspond to $\lambda=1$. A rapid convergence towards
an exponential distribution can already be seen after a time $t \simeq 1$.
}
\label{fig:evolution}
\end{figure}

The evolution of the length distribution in the String model can be obtained as a function of time by explicit numerical integration. 
The results, starting from time $t=0$, are shown in Fig. \ref{fig:evolution}. 

The evolution equation (\ref{evolution-string-model}) can be solved using an exponential ansatz of the form \cite{Krapivsky2010_vola}
\begin{eqnarray}
c_l(t) = (1-a(t))^2 a^{l-1}(t),
\end{eqnarray}
which satisfies the conservation of total number of monomers.
The resulting differential equation for $a(t)$ is $\dot{a}=(1-a)^2-\lambda a$. 
This equation can be easily solved with a monomer-only initial condition, $c_l=\delta_{l,1}$, 
which translates into the condition $a(0)=0$.
Unfortunately, the exponential ansatz cannot be used to describe more general initial conditions, 
which cannot be accounted for by such a simple $l$-independent condition on $a(0)$. 
For the monomer-only initial condition, the following explicit solution is obtained:
\begin{eqnarray}
\label{a(t)}
a(t)&=&\frac{ 1-\exp{\{-(\alpha_{+}-\alpha_{-})t\}} }{ \alpha_{+}-\alpha_{-}\exp{\{-(\alpha_{+}-\alpha_{-})t\}} } \\
&& \alpha_{\pm} \equiv 1+\frac{\lambda}{2} \pm \sqrt{\lambda+\frac{\lambda^2}{4}},
\end{eqnarray}
where the two roots $\alpha_-$ and $\alpha_+$ are related by $\alpha_- \alpha_+=1$.
At long times, {the RHS of} Eq.~\eqref{a(t)} tends towards $1/\alpha_+=\alpha_-=\beta$, so that the
system approaches the equilibrium distribution  
$c_{l}(\infty)=c_{l}^{eq}=(1-\alpha_-)^2 (\alpha_-)^{l-1}$. 

Using (\ref{a(t)}), one can obtain the explicit time evolution of the quantities of interest: 
The total polymer concentration is time-dependent and reads
\begin{eqnarray}
c(t)=\sum_{l \ge 1} c_l(t) = 1-a(t).
\label{C_t}
\end{eqnarray}
The Shannon entropy at the one fluid level, and defined in Eq.~\eqref{Real-Shannon}, is given by
\begin{eqnarray}
S_{Sh}(t)=-\ln{(1-a(t))}-\frac{a(t)}{1-a(t)} \ln{a(t)},
\end{eqnarray}
and reaches its equilibrium value $S_{Sh}(\infty)=-\ln{(1-\alpha_-)}-(\alpha_-/(1-\alpha_-))\ln{\alpha_-}$ for long times.
Its rate of change is
\be
\dot{S}_{Sh}=-a(t) \ln{a(t)}+\lambda\frac{a(t)}{(1-a(t))^2} \ln{a(t)},  \\
\ee
while the entropy production rate given by Eq. \eqref{extensive 2nd law} is
\be
\Sigma = R \left( 2\ln \frac{1-a(t)}{1-\alpha_{-}}-\ln\frac{a(t)}{\alpha_{-}} \right) 
\left( (1-a(t))^2-\lambda a(t) \right).
\ee

\begin{figure}[t!]
\includegraphics[width=0.4\linewidth]{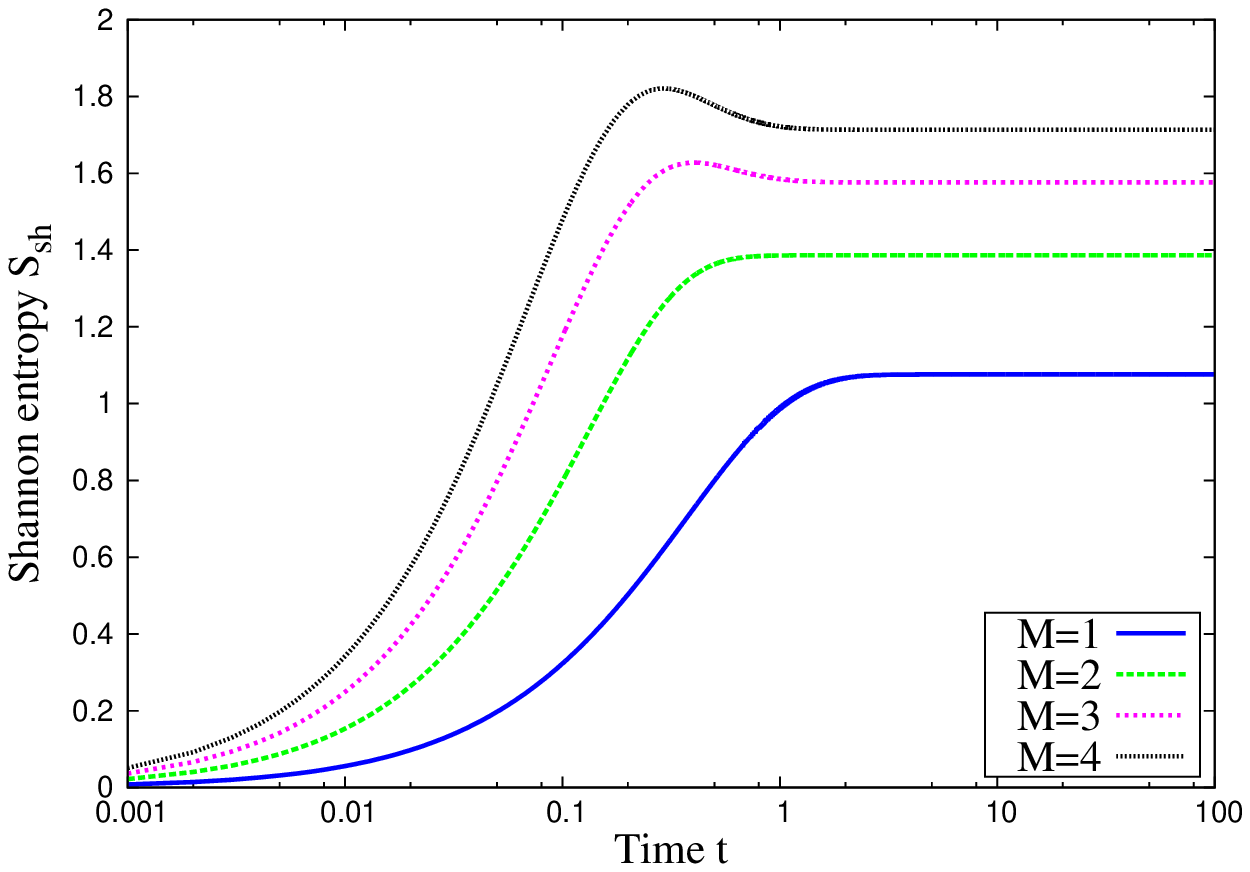}
\includegraphics[width=0.4\linewidth]{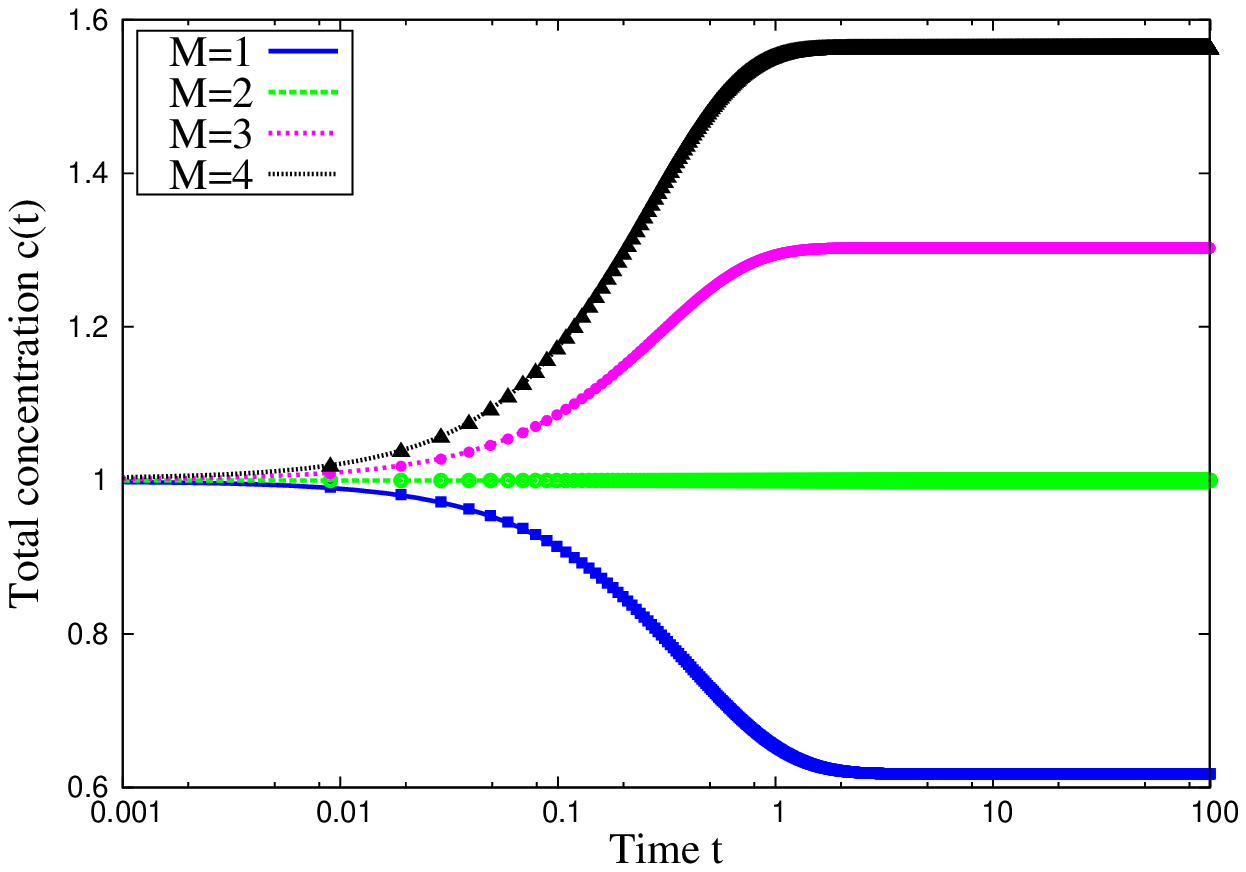}
\caption{\label{fig:Shannon}
Left: Shannon entropy of the one fluid version of the String model, $S_{Sh}$, as a function of time. 
The various curves represent different initial conditions of the form $c_l(t=0)=\delta_{lM}$, 
for different total monomer concentration $M=1,2,3$ and $4$. 
Right: Total polymer concentration $c(t)$ for the same initial conditions, 
as found by numerical integration of kinetic equations (solid line) or 
from an exact expression using Eq. \eqref{explicit C} (symbols). 
Both figures have been done with the same rates as Fig.~\ref{fig:evolution}.
}
\end{figure}

For completeness, we also discuss an approach using generating functions to study the String model 
without resorting to the exponential ansatz which is restricted to the monomer-only initial condition.
The full dynamics remains nevertheless complex to solve within this approach. 
Introducing the generating function $\g(z,t)=\sum_{l \ge 1} c_l(t) z^l$
and using Eq.~\eqref{evolution-string-model}, we obtain the dynamical equation 
\be
\frac{\partial \g}{\partial t}=\g^2 - 2 c \g + 2 \lambda \sum_l \sum_{j>l} c_j z^l - \lambda \left( z \frac{\partial \g}{\partial z} - \g 
\right),
\ee
which can be simplified since for $|z|<1$, $\sum_l \sum_{j>l} c_j z^l= (cz- \g)/(1-z)$, so that 
\be
\frac{\partial \g}{\partial t}=\g^2 - 2 c \g + 2 \lambda \frac{c z - \g}{1 - z} - \lambda \left( z \frac{\partial \g}{\partial z} - \g 
\right).
\label{diff_eq g}
\ee
The stationary solution of this differential equation, i.e. the solution of $\partial \g / \partial t=0$ has the following form:
\be
\g^{eq}(z) \equiv \g(z,t \rightarrow \infty)= \frac{\lambda c^{eq} z}{c^{eq} (1-z) + \lambda},
\label{final sol}
\ee
which satisfies $\g(1,t \rightarrow \infty)=\sum_{l\ge 1}c_l=c^{eq}$ and $\g'(1,t \rightarrow \infty)=c^{eq}(\lambda +c^{eq})/\lambda$. 
Since $M=\sum_{l\ge 1} l c_l=\g'(1,t \rightarrow \infty)=1$, one recovers the equilibrium 
state obtained before, namely $c^{eq}=1- \beta$, with $\beta$ given by Eq.~\eqref{beta}. 
For an arbitrary $M$, one deduces from Eq. \eqref{final sol} that the equilibrium polymer 
length distribution is $c_l^{eq}=\lambda \alpha^l$, where $\alpha=c^{eq}/(\lambda+c^{eq})$. 
With our choice of initial condition of the form $c_l(0)=\delta_{lM}$, we have 
$\sum_{l\ge 1}lc_l=M$. Therefore, the equilibrium solution is
\begin{align}
c_l^{eq} = \frac{(c^{eq})^2}{M}\bigg(1-\frac{c^{eq}}{M}\bigg)^{l-1}.
\label{dist_string}
\end{align}

Besides the stationary solution, it is also possible to obtain analytically the 
evolution of the total concentration $c(t)$. To show this, we take the limit
$z \rightarrow 1$ in Eq. \eqref{diff_eq g}. Using l'Hospital rule, we obtain
\be
\dot{c}=-c^2 + \lambda (M - c).
\label{diff_eq C}
\ee
The explicit solution of Eq.~\eqref{diff_eq C} for $c(0)=1$, as imposed by our choice of initial conditions of the form 
$c_l(t=0)=\delta_{lM}$ is
\be
c(t) = \frac{\Delta}{2} \tanh \left( \frac{t}{2} \Delta 
+ \arctanh \frac{\lambda+2}{\Delta} \right) - \frac{\lambda}{2}. \label{explicit C}
\ee
where $\Delta = \sqrt{\lambda (\lambda+4M)}$.
One can verify that Eq.~\eqref{C_t} is recovered for the monomer only initial condition, $M=1$, as expected. 
Furthermore, one recovers that $c(t)$ tends towards $c^{eq}=(\Delta-\lambda)/2$ as $t\to \infty$, 
which is the equilibrium concentration entering in Eqs. \eqref{final sol} and \eqref{dist_string}.

{Incidentally, one may wonder whether this behavior of the total concentration agrees with the predictions 
of Ref. \cite{Oosawa} regarding the notion of critical concentration in reversible polymerization.
This is indeed the case: if one evaluates the concentration of monomers $c_1$ as a function of $M$, with Eq (\ref{dist_string}) 
and eliminating $c^{eq}$ using the above expression, one finds a function of $M$ which first increases rapidly and then
reaches a plateau for $M \geq \lambda$. Naturally, this is not a sharp transition but rather a cross-over between two regimes.
One can also look at the average length of the polymer $M/c^{eq}$ which increases significantly when $M$ becomes larger than $\lambda$.
Both features indicate that $\lambda$ represents the critical concentration of this model \cite{Oosawa}.}

As shown in the right part of Fig. \ref{fig:Shannon}, which has been obtained by explicit numerical integration, 
the polymer concentration $c(t)$ either decreases as a function of time for the monomer only initial 
condition ($M=1$) or increases as a function of time when the initial condition corresponds to polymers of 
length 3 or above ($M \ge 3$), while it remains constant for the case of dimers ($M=2$). 
Intuitively, when the initial condition is monomer-only ($M=1$), there is mainly aggregation of monomers, 
so that the net concentration must decrease with time. On the other hand, if the initial solution consists 
of long polymers ($M>2$), the probability of fragmentation is higher than that of aggregation. As a result, 
the total concentration must increase with time. For dimer-only initial condition ($M=2$), the probabilities 
of fragmentation is same as that of aggregation, so that the net concentration stays constant.
As a result of this time dependence of $c(t)$, we also see in the left part of Fig. \ref{fig:Shannon}, 
that the Shannon entropy $S_{Sh}$ does not always increase monotonically as a function of time. 
It does so for $M \le 2$ but not for $M >2$, where it presents an overshoot before reaching its equilibrium value. 
Such an overshoot reveals that the Shannon entropy is not a Lyapunov function $L$ as discussed in the previous section.

\subsection{Two-fluid version of the String model}

One of the main difference between the two fluid approach as compared with the previous one with a single
fluid, is the existence of the local detailed balance condition, namely Eq.~\eqref{LDB}, which connects
the rate constants to the difference of standard chemical potentials. Further, the specific form of the standard chemical potentials enters in 
the equilibrium length distribution of the polymers and in the kinetics of the self-assembly process.

For instance, if the polymers self-assemble linearly, the standard chemical potential of a polymer 
of length $l$, $\mu_l^0$ for $l \ge 1$, may be written as $\mu_l^0=-(l-1) \alpha R T$, where $\alpha RT$ 
represents the bond energy between two monomers \cite{Israelachvili1992}. For such a model, a reaction is neutral from the point of view of chemical potentials, i.e. when $\mu_{n+m}^0=\mu_n^0+\mu_m^0$, when the polymer chain is sufficiently long so that $\mu_l^0\simeq -l\alpha RT$.
From Eq.~\eqref{LDB}, it follows that 
\be
RT \ln \frac{k_{nm}^+C^{eq}}{k_{nm}^-}= - (\mu^0_{n+m}-\mu^0_n-\mu^0_m)= \alpha RT,
\ee
which then implies a relation between the parameter $\lambda$ defined earlier as the ratio of the rate 
constants (assumed constant) and the parameter $\alpha$, namely {$\lambda = e^{-\alpha}C^{eq}$}.

Since we introduced $\alpha RT$ as bond energy between two consecutive monomers, the simplest choice is to assume 
that $\mu_l^0$ leads to a molar enthalphy $h_l^0=-(l-1) \alpha R T$, and a molar entropy $s_l^0$ which is assumed to be negligible with respect to the enthalpy part due to bond formation.
With this choice, one finds the following contribution of the polymer to the enthalpy:
\be \label{EnthalpyFct1}
H_1 = -\alpha R T \sum_{l\ge 1} y_l (l-1) {= -\alpha R T (M-c(t))}.
\ee
As discussed previously, at the two fluid level, this should be complemented by the contribution 
of the solvent to obtain the enthalpy $H$.
Similarly, the system entropy $S$, which contains both contributions, is
\be \label{Shannon1}
S = -R \sum_{l\ge 0} y_l \ln{y_l}.
\ee

\begin{figure}[!h]
  \centering
  \includegraphics[width=0.4\linewidth]{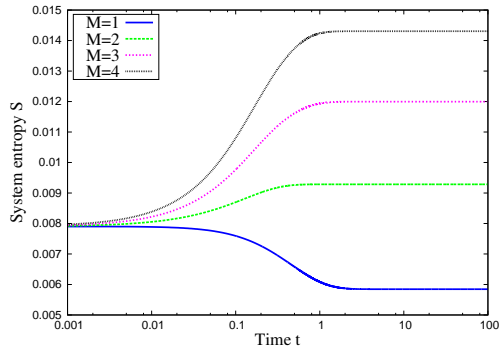}
\caption{\label{fig2} System entropy $S$ for the two-fluid version of the String model as a function of time. 
The various curves represent different initial conditions as in Fig. \ref{fig:Shannon}.
The solvent concentration is $c_0=1000$
and the transition rates are the same as in Fig.~\ref{fig:evolution}. 
}
\end{figure}

In Fig.~\ref{fig2}, we show this entropy function as a function of time for different initial conditions in units of $R$. 
At $t=0$, it equals approximately $(1+\ln c_0)/c_0$, and it converges towards the equilibrium value of the entropy at large times. 
For $M > 1$, the entropy increases monotonically whereas we see that it decreases for $M=1$. As in the case of the one fluid model, 
this decrease is not inconsistent since the entropy is not the Lyapunov function. 
We note that the non-monotonicity that was present in the one-fluid model in 
Fig. \ref{fig:Shannon} for $M >1$ is absent in the two-fluid case. 

If the monomers were to assemble in the polymer in a different way, for instance in the form of disks instead of linear chains, 
the standard chemical potentials would be different. In such a case, under similar assumptions as above, these chemical potentials 
would be of the form $\mu_l^0=-(l-\sqrt{l}) \alpha R T$, where $\alpha$ is again some constant characteristic of the monomer-monomer
and monomer-solvent interaction \cite{Israelachvili1992}. The term in $\sqrt{l}$ represents the contribution of the surface energy 
of the cluster of size $l$. This term necessarily implies that the rate constants $k_{nm}^+$, $k_{nm}^-$ must 
depend on $n$ and $m$ in order to satisfy the local detailed balance condition Eq.~\ref{LDB}. 
For such a case, the above derivation of a simple exponential for the equilibrium distribution 
would no longer hold and both the equilibrium and the dynamics will would be more complex. 

\section{Reversible polymerization model with two conservation laws}
\label{sec:General 2 conservation laws}

As done previously in section \ref{sec:General thermo A} for a reversible polymerization model with only one conservation law, 
namely the total monomer concentration $M$, we now carry out a similar analysis for a different class of models with two 
conservation laws, namely $M$ and the total concentration of polymers or clusters, $c$.  
Clearly, the latter quantity varies in time in the String model because some exchange process 
in Eq.~\eqref{reversible poly general n} produce clusters of zero length for some $n$ or $m$.
In order to construct a model, which conserves the total number of clusters, one needs to forbid such transitions. 
One simple way to achieve this is to consider the kinetic model
\begin{equation}
[n]+[m] \stackrel[k_{nm}]{}{\rightarrow} [n+1] + [m-1],
\label{eq:DPE1}
\end{equation}
with the condition $n \ge 1$ and $m \ge 2$, where the latter inequality precisely prevents the forbidden transitions.
It is easy to check that now the total concentration of the polymers, $c = \sum_l c_l$, as well as the total 
monomer concentration $M = \sum_l lc_l$, remain constant in time. 
Another important observation is that this model is fully reversible even if we do not indicate backward reactions explicitly. 
Indeed, it would be redundant to do so, since backward reactions are already 
included in the forward reactions via an appropriate choice of the indexes $(n,m)$. 
As done in the previous section, we first present a general proof of convergence to equilibrium and then we make contact 
with thermodynamics by introducing chemical potentials in dilute solutions. 

The equation for the rate of change of concentration for polymer size distribution is
\begin{align}
  \dot c_l &=\Theta(l-2)\sum_{n=1}^\infty [k_{l-1,n+1} c_{n+1}c_{l-1} - k_{nl} c_nc_l] + \sum_{m=2}^\infty[k_{m-1,l+1} c_{l+1}c_{m-1} - k_{lm} c_lc_m],
  \label{cldot}
\end{align}
where the Heaviside function $\Theta(l-2)$ equals 1 for $l \ge 2$, and is zero otherwise. 

Assuming again elementary reactions, the entropy production rate $\Sigma$ is 
\begin{align}
\Sigma = \frac{R}{2}\sum_{n\ge 1,m\ge 2}[k_{nm} c_nc_m-k_{m-1,n+1} c_{n+1}c_{m-1}] \ln \frac{k_{nm}c_nc_m}{k_{m-1,n+1}c_{n+1}c_{m-1}} \geq 0,
\end{align}
which vanishes when the detailed balance condition holds, {\it i.e.} in equilibrium:
\begin{align}
k_{nm}c_n^{eq}c_m^{eq} = k_{m-1,n+1} c_{n+1}^{eq}c_{m-1}^{eq} .
\label{DB2}
\end{align}
Following a procedure similar to that of section \ref{sec:General thermo A}, one can show that, 
since $c$ is constant, the relative entropy between the distribution $x_l$ and $x_l^{eq}$
\begin{align}
L = R \sum_l c_l \ln \frac{c_l}{c^{eq}_l} = R c \sum_l x_l \ln \frac{x_l}{x^{eq}_l},
\label{L:2}
\end{align}
is a Lyapunov function. Indeed, this quantity is convex, non-negative (by the inequality $\ln x < x-1$), 
and a monotonically decreasing function vanishing at equilibrium. This latter property follows from 
\be 
\frac{dL}{dt}= R \sum_l \dot c_l\ln\frac{c_l}{c^{eq}_l}, 
\ee
which using Eq. \eqref{cldot} and the detailed balance condition of Eq. \eqref{DB2} gives 
\be
\frac{dL}{dt}= R \sum_{n \ge 1, l \ge 2} k_{nl} c_n c_l \ln \frac{k_{l-1,n+1} c_{n+1} c_{l-1}}{k_{nl} c_n c_l}.
\ee
After symmetrizing this sum, we recover that $\Sigma=-dL/dt \geq 0$. This result is equivalent to Eq.~\eqref{extensive 2nd law} 
in presence of the additional conservation law $\dot{c}=0$. This system will therefore relax to a unique 
equilibrium state, where $\Sigma$ vanishes.

We now turn to the two fluid version of the model. As in section \ref{subsec:1conser_2fluids}, for the two fluids 
model the molar fraction of a polymer of length $l$ is $y_l(t)=c_l(t)/C$, where $C=c+c_0$ is again a constant. 
The change in chemical potential during the reaction (\ref{eq:DPE1}) is given by
\begin{align}
  \Delta\mu &= \mu_{n+1}+\mu_{m-1}-\mu_n-\mu_m \nn\\
  &= \mu^0_{n+1}+\mu^0_{m-1}-\mu^0_n-\mu^0_m + RT\ln\frac{y_{n+1}y_{m-1}}{y_ny_m}.
\end{align}
Since at equilibrium $\Delta \mu=0$, using (\ref{DB2}), we get that
\begin{align}
  \mu_n^0+\mu_m^0-\mu^0_{n+1}-\mu^0_{m-1} = RT\ln\frac{y^{eq}_{n+1}y^{eq}_{m-1}}{y^{eq}_ny^{eq}_m} =RT\ln \frac{k_{nm}}{k_{m-1,n+1}}.
\end{align}

The enthalpy change (\ref{EnthalpyFct}) can be written as
\begin{align}
  \frac{d {\cal H}}{dt} &= C\frac{d}{dt} \sum_l y_l h^0_l = \sum_l \dot c_lh^0_l \nn\\
  &= \sum_{n\ge 1,m\ge 2}(k_{m-1,n+1}c_{n+1}c_{m-1}-k_{nm}c_nc_m)(h^0_n+h^0_m) \nn\\
  &= \sum_{n\ge 1,m\ge 2}(k_{nm}c_{n}c_{m} - k_{m-1,n+1}c_{m-1}c_{n+1})(h^0_{m-1}+h^0_{n+1}) \nn\\
  &= \frac{1}{2}\sum_{n\ge 1,m\ge 2}(k_{nm}c_nc_m - k_{m-1,n+1}c_{n+1}c_{m-1})(h^0_{m-1}+h^0_{n+1}-h^0_n-h^0_m)
\end{align}
and the entropy change (\ref{Shannon}) as
\begin{align}
  \frac{d{\cal S}}{dt} &= C\frac{d}{dt}\sum_{l\ge 0} y_l(s^0_l-R\ln y_l) = \sum_{l\ge 0}\dot c_l(s^0_l-R\ln c_l)\nn\\
  &= \sum_{n\ge 1,m\ge 2}(k_{nm}c_nc_m - k_{m-1,n+1}c_{n+1}c_{m-1})\left(s^0_{m-1}+s^0_{n+1}-s^0_n-s^0_m+R\ln\left[\frac{y_ny_m}{y_{n+1}y_{m-1}}\right]\right).
\end{align}
Since the entropy production can be rewritten as
\begin{align}
  T\Sigma &= \frac{1}{2}RT\sum_{n\ge 1,m\ge 2}[k_{nm}c_nc_m - k_{m-1,n+1}c_{n+1}c_{m-1}]\ln\frac{k_{nm}c_nc_m}{k_{m-1,n+1}c_{n+1}c_{m-1}} \nn\\
  &= \frac{1}{2}\sum_{n\ge 1,m\ge 2}[k_{nm}c_nc_m - k_{m-1,n+1}c_{n+1}c_{m-1}]\left[\mu^0_n+\mu^0_m-\mu^0_{n+1}-\mu^0_{m-1}+RT\ln\frac{y_ny_m}{y_{n+1}y_{m-1}}\right] \nn\\
  &= \frac{1}{2}\sum_{n\ge 1,m\ge 2}(k_{nm}c_nc_m - k_{m-1,n+1}c_{n+1}c_{m-1})(\mu_n+\mu_m-\mu_{n+1}-\mu_{m-1}),
\end{align}
we can express it, as in (\ref{EntBalance2}), as
\begin{align}
T \Sigma = -\frac{d}{dt} ({\cal H}-T{\cal S}) 
= -\frac{d {\cal G}}{dt} 
= - R \frac{d}{dt} \big( C \sum_{l\ge 0} y_l  \ln \frac{y_l}{y_l^{eq}} ) 
= -\frac{d L}{dt} \geq 0.
\end{align}
To summarize, we recover exactly the same results as in section II, 
provided we treat the total polymer concentration $c$ as a constant.

\section{Application to the kinetics of glucanotransferases DPE1 and DPE2}
\label{sec:DPE1-DPE2}

In this section, we consider the polymerization of glycans by two enzymes studied by 
Kartal et al. \cite{Kartal2011}, namely the glucanotransferases DPE1 and DPE2.
We show how to construct dynamical models that are compatible with the equilibrium polymer length distributions 
that they found and we study the dynamics of the Shannon entropy for various initial conditions.

\subsection{Kinetics of glucanotransferases DPE1}

Let us assume that the initial condition is not purely made of monomers, since the solution of Ref. \cite{Kartal2011}
becomes singular in that limit (see Eq. (4) on P3 of Ref. \cite{Kartal2011} when the parameter $DP_{in}=1$ for instance), 
and let us construct an appropriate dynamics, choosing for simplicity constant rates $k_{nm} = \kappa$ independent of $n$ and $m$. 

Using Eq. \eqref{cldot}, we have for $l \ge 2$,  
\be
\frac{dc_l}{dt}= \kappa [c \left( c_{l+1}+c_{l-1}-2c_l \right) - c_1 c_{l-1} + c_1 c_l],
\ee
where the {second term} forbids transitions from $[1]+[l-1]$ to $[0]+[l]$, while
the last term forbids transitions from $[1]+[l]$ to $[0]+[l+1]$.
Similarly, for $c_1$, the evolution is
\be
\frac{dc_1}{dt}= \kappa [c \left(c_2 - c_1 \right) + c_1^2 ].
\ee
It is straightforward to verify that this dynamics has two conservation laws,
namely $\sum_{l \ge 1} c_l=c$ and $\sum_{l \ge 1} l c_l=M$. 

Introducing the generating function $\g(z,t)=\sum_{l \ge 1} c_l(t) z^l$ as in section 
\ref{sec:Application string model} leads once again to a set of equations for the dynamics 
which unfortunately can not be solved analytically. However, it enables us to find an explicit 
solution for the equilibrium state:
\be
\g^{eq}(z) \equiv \g(z,t \rightarrow \infty)= \frac{c^2 z}{M - M z + c z},
\label{G_inf}
\ee
which means that the size distribution $c_l$ tends towards the following 
equilibrium distribution for $l \ge 1$: 
\be
\label{Eq-dist}
c^{eq}_l= \frac{c^2}{M} \left( 1 - \frac{c}{M} \right)^{l-1},
\ee
where the total polymer concentration $c$ is now fixed by the initial condition. 
We note that the form of the equilibrium distribution is the same as that of the 
String model, but $c^{eq}$ in Eq. \eqref{dist_string} is different from $c(t=0)$, 
whereas in the present DPE1 model they are the same.
Our equilibrium solution (\ref{Eq-dist}) also matches that of Kartal et al. found for 
the polymerization of glycans by glucanotransferases DPE1 \cite{Kartal2011}. 
In this reference, the authors use the polymer fractions $x_l$, where $l$ stands for the number of 
linkages in one cluster, rather than our cluster distribution $c_l$. They are related by $x_l=c_{l+1}/c$. 
Their conservation laws therefore read $\sum_{l \ge 0} x_l=1$ and $\sum_{l \ge 0} l x_l=DP_{ini}-1$, 
where $DP_{ini}$ stands for the initial degree of polymerization. The latter is related to $c$ by
$DP_{ini}=M/c$, and the relation $1-c/M=e^{-\beta}$ matches Eq. (4) of Kartal et al.

\begin{figure}[!h]
\centering
\includegraphics[width=0.4\linewidth]{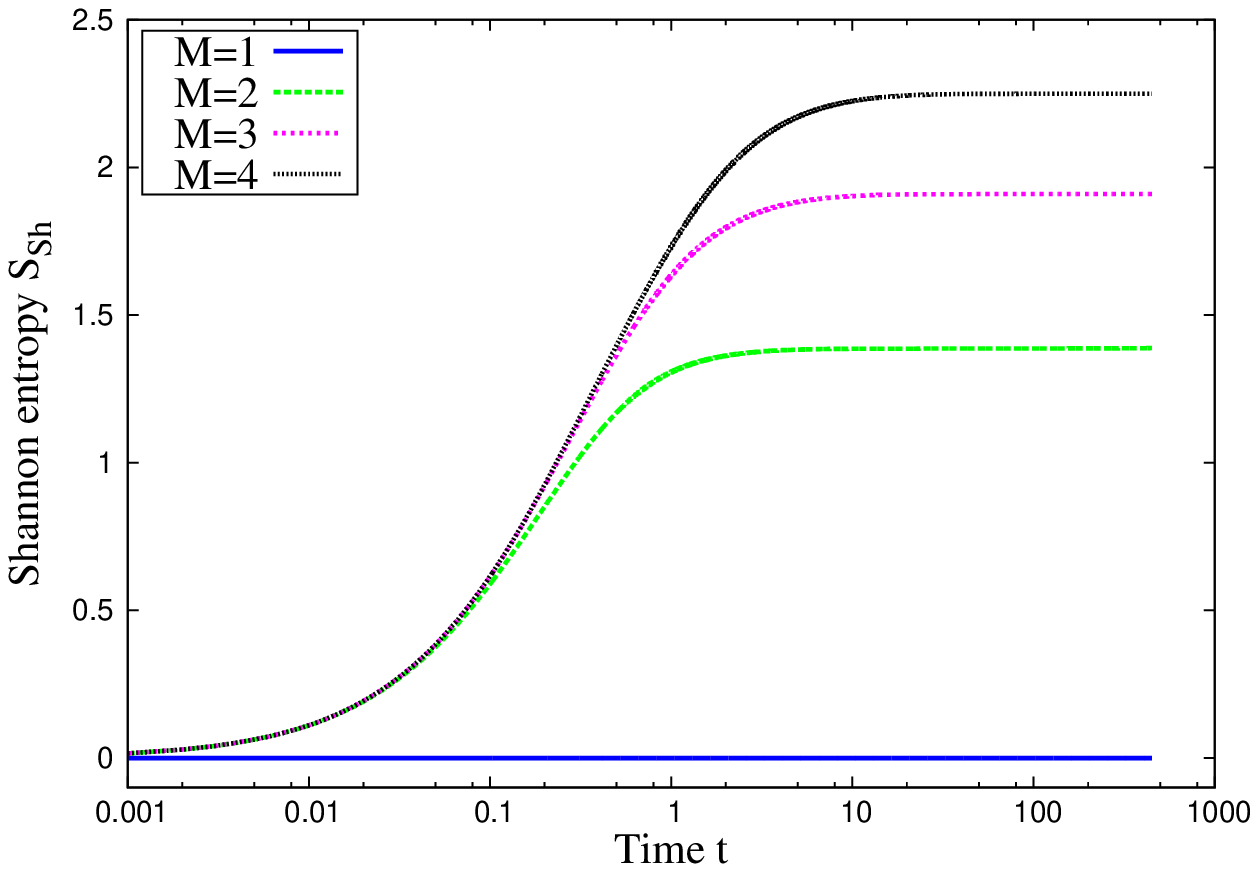}
\includegraphics[width=0.4\linewidth]{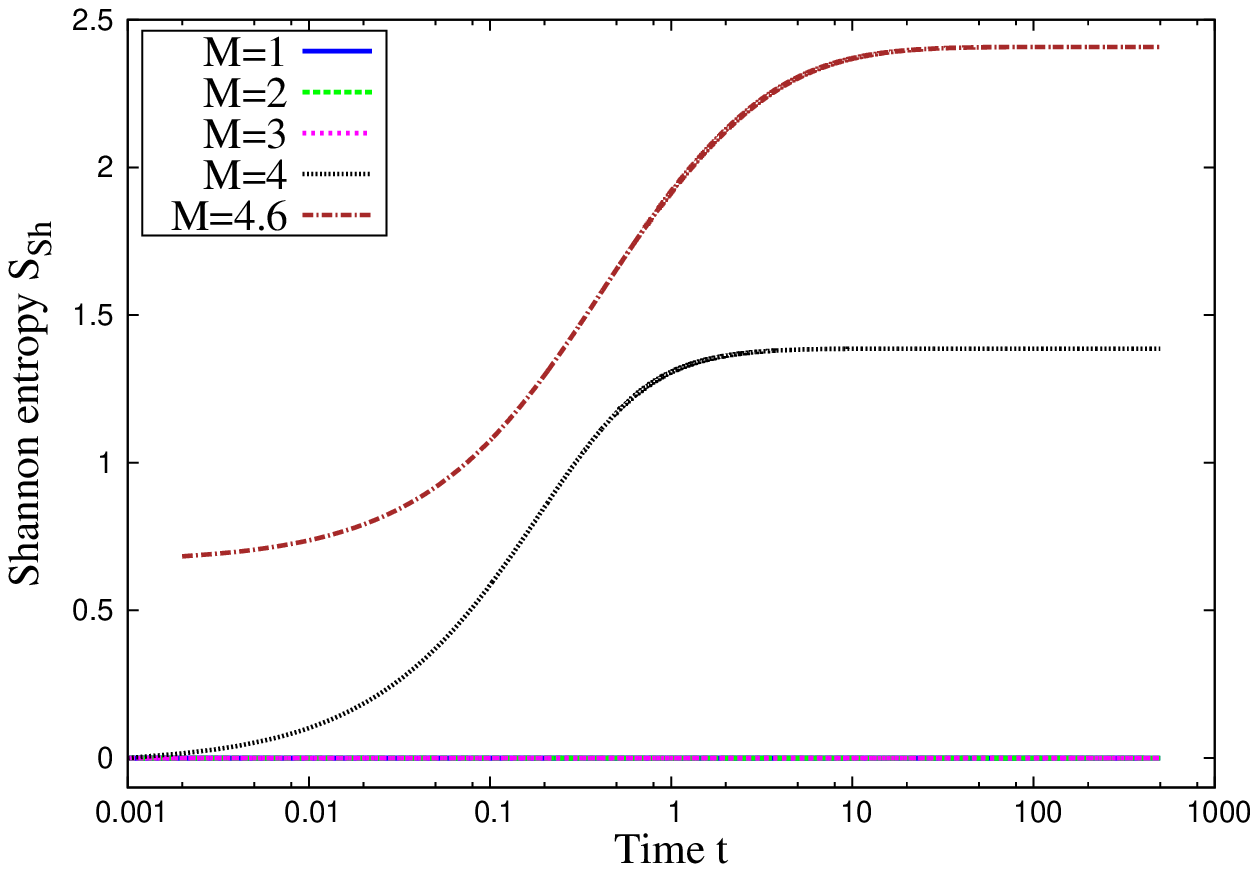}
\caption{\label{fig3}
Left (resp. right): Evolution of the Shannon entropy $S_{Sh}$ for the one fluid DPE1 (resp. DPE2) model 
with time for different initial conditions. For both plots, the curves for $M=1,2,3$ and $4$ correspond 
to initial conditions of the form $c_l(0)=\delta_{lM}$.
For the DPE2 plot on the right, $M=4.6$ corresponds to an initial condition of the form 
$c_l(0)=0.4 \delta_{l1}+0.6 \delta_{l7}$.  
Note that when $M=1$ for DPE1 and $M=1,2$ and $3$ for DPE2, the  
Shannon entropy is zero since no evolution is possible from the initial condition. 
We have chosen the constant rate $\kappa=1$.}
\end{figure}

The Shannon entropy, Eq.~\eqref{Real-Shannon}, at equilibrium and in $R$ units, reads   
\be
S_{Sh}(t \rightarrow \infty)=\frac{M}{c} \ln \frac{M}{c} - \left( \frac{M}{c}-1 \right) \ln \left( \frac{M}{c}-1 \right),
\ee
and has the standard form of a mixing entropy. Note that the case of monomer-only
initial condition, namely $c=M=1$, is singular since no evolution is possible from this
initial condition according to the present dynamics. In this case, the Shannon entropy 
stays at zero, for all times $t$, whereas for other initial conditions it 
increases monotonically as shown in Fig. \ref{fig3}. 
 
\subsection{Application to the kinetics of glucanotransferase DPE2}

As discussed by Kartal et al. \cite{Kartal2011}, the enzyme DPE2 introduces an additional constraint with 
respect to the enzyme DPE1. This additional constraint, {which imposes a conservation of 
the total number of monomers and dimers}, reads in our notations
$c_1 + c_2 = pc$, where $p$ depends on the initial conserved total number of molecules of
maltose (corresponding to $c_2$) and glucose (corresponding to $c_1$).
This additional constraint requires a modification of the dynamical evolution equations. 
We propose the following modification:
\ba
\frac{dc_1}{dt} &=& \kappa [c \left(c_2 - c_1 \right) + c_1^2 - c_2^2 + c_1 c_3], \\
\frac{dc_2}{dt} &=& - \frac{dc_1}{dt}, \\
\frac{dc_3}{dt} &=& \kappa [c (c_4 -c_3) - c_2 c_4 + c_1 c_3 + c_3^2], 
\ea
and for $l \ge 4$,
\ba
\frac{dc_l}{dt} &=& \kappa [c \left( c_{l+1}+c_{l-1}-2c_l \right) - c_1 c_{l-1} - c_2 c_{l+1} \nonumber \\
&&- c_3 c_{l-1} + (c_1+c_2+c_3) c_l],
\ea

As in the case of DPE1, one can solve the stationary state of this equation by means 
of generating functions. One obtains the following stationary generating function:
\be
\label{EqDistDPE2}
\g^{eq}(z) \equiv \g(z,t \rightarrow \infty)= \frac{ \left[ c_1 c_2 - c c_1 - z^2 
\left( c_1 c_2 - c c_2 + c c_3\right) \right] z}{c(z-1) -c_1 z + c_2 - c_3 z}.
\ee
One obtains from this $c_1=pc/(1+f)$ and $c_2=c_1 f=fpc/(1+f)$ with $f=(c - c_1 - c_3)/(c - c_2)$, 
and for $l \ge 3$, $c_l = c_3 f^{l-3}$ with $c_3 = (1-p) (1-f) c$.
In other words, for DPE2, the equilibrium distribution is again exponential but only for length $l \ge 3$, 
for $l<3$ the ratio of $c_2/c_1$ for instance does not match the ratio $c_{l+1}/c_l$ for $l \ge 3$.
The quantity $f$ can be written in terms of $p$ and $c$ only 
\begin{eqnarray}
M - 2 c (1- \frac{1}{2} p) = p c \frac{f}{1+f} + c (1-p) \frac{1}{1-f},
\end{eqnarray}
which matches Eq.~(S57) obtained by Kartal et al. \cite{Kartal2011}.
Therefore, the equilibrium state (\ref{EqDistDPE2}) is the same as that discussed in this reference.

We have thus proposed dynamical models reproducing the equilibrium distribution of glycans in presence of DPE1 or DPE2. 
The difference between both situations is that DPE1 has two conservation laws, namely that of $M$ and of $c$,
while DPE2 has a third one corresponding to that of $p$. As a result, there are more initial conditions 
of the type $c_l(0)=\delta_{lM}$ from which no evolution is possible in DPE2 ($M \le 3$) as compared to DPE1. 
When this happens, $S_{Sh}=0$ as shown in Fig. \ref{fig3}. 
While this forbids initial conditions of pure dimers for instance, no such constraint exists for mixtures. 
For instance, an initial mixture of 40:60 of maltose and maltoheptaose considered in \cite{Kartal2011}, 
corresponding to $c_l(0)=0.4 \delta_{l1}+0.6 \delta_{l7}$, has $p=0.4$ and $M=4.6$, and 
evolves according to DPE2 dynamics as shown in Fig. \ref{fig3}, while an initial solution 
of pure maltose would not. 

The limiting value of the Shannon entropy at long times can be obtained analytically as a function of $f$ 
and $p$ for any initial conditions, but the expression is lengthy and will not be given here. 
We have checked that it reproduces the correct value of the plateaux in Fig. \ref{fig3}.  

\section{Conclusion}

In this paper, we have considered two classic models for reversible polymerization in closed systems following 
the mass-action law, one preserving the total polymer concentration and the other one not. In both cases, the entropy 
production can be written as the time derivative of a Lyapunov function which guarantees the relaxation of 
any initial condition to a unique equilibrium satisfying detailed balance. As such, these models could 
also describe non-chemical systems undergoing an aggregation-fragmentation dynamics.    

When considering the polymerization dynamics in dilute solutions, we have shown that a consistent 
nonequilibrium thermodynamics can be established for both models. We find that entropy 
production is minus the time derivative of the nonequilibrium free energy of the system, which is  
a Lyapunov function and takes the form of a Kullback-Leibler divergence between 
the nonequilibrium and the equilibrium distribution of polymer length.
A related result was found for the cyclical work performed by chemical machines feeding on polymers in Ref. \cite{Smith08}.    
Similar relations expressing the entropy production as a Kullback-Leibler divergence between the nonequilibrium and 
equilibrium distributions have also been found or used 
in many studies on Stochastic thermodynamics \cite{JarzynskiEPL09, EspoVdB_EPL_11,Tusch2014}.

As an application of reversible polymerization models which do not preserve the total polymer concentration, 
we have studied the String model. In this model, the rates of aggregation and fragmentation are constants,
which leads to an exponential equilibrium distribution of polymer length. 
At the one-fluid level, we have observed that the Shannon entropy is non-monotonic, which is allowed since 
it differs from the Lyapunov function. 
At the two-fluid level where there is a proper
nonequilibrium thermodynamics, no such non-monotonicity arises. 

As an application of reversible polymerization models preserving the total polymer concentration 
in addition to the total number of monomers, we have studied two specific examples named 
DPE1 or DPE2 after Ref. \cite{Kartal2011}. We have shown how to construct dynamics which converge at long times to 
the expected form  and we have discussed the time evolution of the Shannon entropy at the one-fluid level.
{In all the cases, we have been able to find the form of the stationary distribution, by applying the method of 
generating functions. This method is general and also applicable to situations where the stationary distribution is a nonequilibrium one \cite{Ranjith2009}.}

Key assumptions of our approach are that we disregarded fluctuations, assumed homogeneous and ideal solutions, 
considered closed systems, and we treated the polymerization reactions as elementary. Each of these 
assumptions could in principle be released and the resulting implications analyzed.
Another interesting future direction concerns the study of  
nonequilibrium thermodynamic devices or strategies 
which can be used to engineer a particular polymer distribution
(for instance a monodisperse one) starting from an initial polydisperse one (an exponential one for instance).  
 
\section*{Acknowledgments}

We acknowledge stimulating discussions with Oliver Ebenhöh, Riccardo Rao and Alexander Skupin. 
{We would like to also acknowledge P. Gaspard for a critical reading of this work and for insightful comments.}
M. E. is supported by the National Research Fund, Luxembourg in the frame of project 
FNR/A11/02 and S. L. by the Region \^Ile-de-France thanks to the ISC-PIF.

\vskip 10pt

%

\end{document}